Title : Effects and limitations of a nucleobase-driven backmapping procedure for nucleic acids using steered Molecular Dynamics.

Authors : Simón Poblete[a,*], Sandro Bottaro[b] and Giovanni Bussi[a]

[a] Molecular and Statistical Biophysics Group, Scuola Internazionale Superiore di Studi Avanzati, 265, Via Bonomea I-34136 Trieste, Italy.

[b] Structural Biology and NMR Laboratory, Department of Biology, University of Copenhagen, Ole Maaloes Vej 5, 2200 Copenhagen N, Denmark

spoblete@sissa.it (Simón Poblete)
sandro.bottaro@bio.ku.dk (Sandro Bottaro)
bussi@sissa.it (Giovanni Bussi)

[*] Corresponding author. Molecular and Statistical Biophysics Group, Scuola Internazionale Superiore di Studi Avanzati, 265, Via Bonomea I-34136 Trieste, Italy.

# Abstract

Coarse-grained models can be of great help to address the problem of structure prediction in nucleic acids. On one hand they can make the prediction more efficient, while on the other hand, they can also help to identify the essential degrees of freedom and interactions for the description of a number of structures. With the aim to provide an all-atom representation in an explicit solvent to the predictions of our SPlit and conQueR (SPQR) coarse-grained model of RNA, we recently introduced a backmapping procedure which enforces the predicted structure into an atomistic one by means of steered Molecular Dynamics. These simulations minimize the $\mathcal{E}$RMSD, a particular metric which deals exclusively with the relative arrangement of nucleobases, between the atomistic representation and the target structure. In this paper, we explore the effects of this approach on the resulting interaction networks and backbone conformations by applying it on a set of fragments using as a target their native structure. We find that the geometry of the target structures can be reliably recovered, with limitations in the regions with unpaired bases such as bulges. In addition, we observe that the folding pathway can also change depending on the parameters used in the definition of the $\mathcal{E}$RMSD and the use of other metrics such as the RMSD.

# Introduction

Despite the computational advantages that coarse-grained (CG) models can provide to the study of biological and soft matter systems [1], a consistent all-atom representation of their results is required in a vast number of cases. The validation of their predictions against data obtained from X-ray or neutron scattering experiments, as well as their use in the generation of equilibrated configurations for all-atom simulations, are common situations of this type found in multiscale modelling.

Usually, a backmapping procedure starts with the determination of the atomistic coordinates from analytical expressions of the positions of the CG sites, or by properly assembling atomistic fragments consistent with the CG mapping. Later on, depending on the complexity of the system and the desired accuracy, the energy of the resulting structure can be minimized by methods such as steepest descent or equilibrated through short MD simulations [2]. This step can be performed under position restraints in order to keep the correspondence between both atomistic and CG representations. Several implementations of these approaches have been successfully applied in the study of polymer melts [2–8] or proteins, lipids and peptides [9–19], and also been defined in a more systematic way [15]. If the backmapping procedure is efficient, it is also possible to produce entire trajectories with all-atom resolution from CG trajectories [3]. When using CG models for structure prediction, it is a comparably small set of candidate structures which has to be backmapped instead. It is also often desirable to use them in later all-atom simulations, so a relaxed configuration embedded in explicit solvent without clashes should also be present together with a reliable contact map. In the case of nucleic acids, several structure prediction methods provide their own tools for reconstructing all-atom structures as Vfold [20] or HiRe-RNA [21], which also make use of a fragment assembly approach with further solvation and in some cases, a relaxation in all-atom representation.

In this paper, we discuss in detail the nucleobase-centered backmapping procedure introduced in our previous work of the SPlit and conQueR (SPQR) CG model [22]. The method minimizes the $\mathcal{E}$RMSD distance [23] between solvated, all-atom strands and a target CG structure. This distance depends exclusively on the difference between the arrangement of nucleobases of two given RNA structures, so it requires the unambiguous definition of both the position and orientation of each nucleobase, as it is the case of the SPQR and many other CG representations of RNA [21,24–26]. On the other side, backbone atoms are unrestrained, and they accommodate according to the force field. The importance of the directional interactions between nucleobases such as non-canonical base pairs lead to their explicit inclusion in several recent CG models [27,25,28]. It is then of interest to understand what is the relevance of the sole base arrangement in determining the full RNA structure.

For the analysis, we will focus on the effects of the backmapping on the backbone conformations and interaction network. In order to avoid the problems that could emerge from the limitations of a particular CG prediction, we will perform the backmapping on a set of RNA fragments using as a target their respective native structures.

Concretely, the backmapping procedure makes use of steered-Molecular Dynamics [29] which forces the system towards a minimum $\mathcal{E}$RMSD distance between the target and the all-atom representation, thus in the spirit of targeted MD [30]. Similar approaches have also been used in protein folding, using the native contacts as steered variable [31], although not in the context of backmapping. We notice also that the $\mathcal{E}$RMSD distance has been already used as a collective variable for the calculation of the stability of the folded structure of RNA hairpins [32]. The procedure used here is less ambitious in the sense that it does not aim at inducing reversible folding, and is thus significantly less expensive.

The paper begins with the introduction of the employed methodology and the parameters used. Later on, the main results are exposed, concerning the convergence of the procedure and an analysis of the contact map, treating base-base and base-phosphate interactions separately. It follows an analysis of the consistency of the glycosidic bond angles and sugar puckers with the target counterparts of each structure. Two specific cases are studied in more detail: an UUCG tetraloop and a viral RNA pseudoknot. These structures are used for testing different parameters such as the strength of the pulling force used in the steered-MD, and to compare with an alternative approach which minimizes the RMSD [33] of a specific set of atoms between atomistic and target structures. We conclude with a summary of the limitations and advantages of our backmapping procedure according to the results presented here.

## Materials and Methods

We tested the backmapping method on 12 fragments, which are listed in Table 1 with their corresponding number of nucleotides, resolution and description. The steered-MD simulations were performed on atomistic RNA in explicit water (TIP3P water molecules [34], Amber99 force field [35] with parmbsc0 [36] and χOL3 corrections [37]) in a truncated dodecahedral box with Na+ counterions [38] in Gromacs 4.6.7 [39]. The corrections to the potential [36,37] were particularly designed to remove previous artifacts which affect mainly the helical structure of RNA through the α, γ and χ dihedrals on timescales longer than 10 ns. Although this force field was shown not to be able to reproduce the stability of some hairpin loops [32,40], to our knowledge it can be considered as the most tested and reliable available choice. The initial conditions were generated with the make-na server (http://structure.usc.edu/make-na/server.html, August 2017), which correspond to one or two RNA strands as in an A-form, depending on the system. In the case of duplexes, the strands were generated by the same procedure and manually placed on positions separated by a distance around 10 Å. The energy of these conformations was minimized by steepest-descent before and after adding water and ions for 5000 and 4000 steps, with a maximum force of 1.0 kJ mol$^{-1}$nm$^{-1}$. Later on, a short equilibration run was performed for 1 ns. 10 initial conditions were initialized with different seeds, and run for 3 ns using the Stochastic Velocity Rescaling thermostat [41] at 300 K under the restraining force. In general, RNA structural dynamics as obtained from MD simulations is largely independent of the employed concentration of monovalent cations (see e.g. [42]). Thus, matching the ion concentration to experimental values is not expected to improve the results. Also, the effects of the cations on the backbone conformations are not explicitly analyzed here. In fact, it has been observed that even strongly binding divalent ions do not affect the backbone conformations in duplexes [43]. Clearly, the detailed characterization of motifs that explicitly require ion binding to fold is out of the scope of the present work.

$\mathcal{E}$RMSD is a metric which quantifies how different are the internal arrangement of bases between two given structures. Given a three-dimensional RNA structure, we set up a local coordinate system with the origin in the center of mass of the set composed by the C2, C4 and C6 atoms. For each pair of bases $(i,j)$ in a molecule we calculate $r_{ij}$, i.e., the position of the center of base $i$ in the coordinate system of base $j$. In order to account for the anisotropy in base-base interactions, we consider the scaled vector $\tilde{r} = \left(\frac{r_x}{a}, \frac{r_y}{a}, \frac{r_z}{b}\right)$ with $a = 5$ Å and $b = 3$ Å. The $\mathcal{E}$RMSD, between two structures $\alpha$ and $\beta$ is defined as

$$\mathcal{E}\text{RMSD} = \sqrt{\frac{1}{N}\sum_{j \neq k}\left(G(\tilde{r}_{jk}^{\alpha}) - G(\tilde{r}_{jk}^{\beta})\right)^2}$$

Where $G$ is a four-dimensional smooth function such that $\left|G(\tilde{r}^\alpha) - G(\tilde{r}^\beta)\right| \approx \left|\tilde{r}^\alpha - \tilde{r}^\beta\right|$ if $\tilde{r}^\alpha, \tilde{r}^\beta \ll \mathcal{D}_c$ and $\left|G(\tilde{r}^\alpha) - G(\tilde{r}^\beta)\right| = 0$ if $\tilde{r}^\alpha, \tilde{r}^\beta > \mathcal{D}_c$ and $\mathcal{D}_c$ is a suitable unitless cutoff [23].

The ellipsoidal distance used in the calculation of $\mathcal{E}$RMSD requires the sole knowledge of the position of the C2, C4 and C6 atoms of each nucleobase (or a way to define them unambiguously), which are extracted from the PDB files of the target structures. The $\mathcal{E}$RMSD was calculated using a cutoff of $\mathcal{D}_c=3.6$, unless other value is indicated. Note that it differs from the usual RMSD in several aspects: $\mathcal{E}$RMSD depends only on the mutual arrangement of nucleobases, and considers all the pairs separated by an ellipsoidal distance smaller than $\mathcal{D}_c$. On the other hand, RMSD takes into account the displacement of the position of each particle with respect to a reference structure.

The external pulling force was applied using the MOVINGRESTRAINT option in PLUMED [44], which applies a potential of the form

$$U(s(t), t) = \frac{1}{2} \kappa(t) \times s(t)^2$$

where $s(t)$ is the $\mathcal{E}$RMSD with respect to the target structure at time $t$ and

$$\kappa(t) = \begin{cases} 0 & t < t_0 \\ \kappa_{max} \dfrac{(t - t_0)}{t_1 - t_0} & t_0 < t < t_1 \\ \kappa_{max} & t > t_1 \end{cases}$$

with $t_0 = 100$ ps, $t_1 = 2$ ns and $\kappa_{max} = 1000$ kJ/mol. From these simulations, we selected the lowest $\mathcal{E}$RMSD configuration from each of the trajectories for the interaction network analysis, and the lowest 50 for the backbone analysis. The backbone dihedrals were calculated using the Dangle package (http://kinemage.biochem.duke.edu/software/dangle.php, August 2017). The comparison of the sugar puckers is done by comparing the value of the $\delta$ backbone dihedral. We consider that their values are equivalent if they have the same conformation, or if their difference is smaller than a predefined cutoff and if this difference is smaller than the distance to each of the closest reference value of each conformation. The cutoff and the reference values are taken according to the average radius and centroids of the clusters identified in the suite families in the Suitename program (http://kinemage.biochem.duke.edu/software/suitename.php, August 2017). For the glycosidic bond angle $\chi$, the conformations are *anti*, *syn* and *intermediate syn*, consistently with the definition of [45].

The interaction network fidelity (INF) score [46] is used for comparing the contact map of different structures, and is calculated separately for base pairs, stacking and base-phosphate interactions. The structures are annotated using the FR3D package [47].

A simulation is considered to converge when the minimum $\mathcal{E}$RMSD value observed is lower than $\mathcal{E}\text{RMSD}_{min} = 0.8$ with of $\mathcal{D}_c=3.6$, a value suitable for our convergence criteria.

# Results

## Convergence towards target structure

The $\varepsilon$RMSD as a function of time for one of the systems analyzed, the UUCG tetraloop (PDB: 2KOC), is reported in Figure 1a. We observe that it decays quickly approximately 500 ps after the force has been applied. The stem is formed in all the simulations, although the loop region is not always properly folded, as illustrated in Figures 1b and 1c. This mismatch, however, is likely due to the speed of the pulling trajectory and is clearly identifiable from their $\varepsilon$RMSD difference with respect to the native structure.

After selecting the best structures according to the aforementioned criteria, we assess the accuracy of the arrangement between bases and phosphate groups by calculating the INF score in these fragments and their respective native structures. We report in Table 2 the number of initial conditions which converged to the native structure and the results of the average INF for base pairs, stacking, base-phosphate and non-canonical base pairs.

While the INF scores for stacking and base-pairing are almost perfect for all fragments, the non-canonical score is slightly worse. In 1ZIH and 2GDI the low scores correspond to false positives which vanish when the annotation criterion becomes more tolerant, as it is shown with the number in parenthesis annotated with the "nearly" interactions also considered [47]. This implies that the false positives found in our simulations do have a physical meaning. On the other hand, the base-phosphate interactions, which are also the scarcest contacts, exhibit relatively low scores in general. This is mainly due to the lack of constraints on the phosphate groups position and orientation. In many cases, the contacts are simply missing. Nevertheless, in cases such as 2CKY and 2OIU, the low score is due to the exchange between 3BPh and 4BPh interactions [48], which differ only by one hydrogen bond but involve the same face of the base. The tetraloops 1ZIH and 2KOC show different results. Although the base-phosphate contact is present sometimes in the first case, the latter seems to be quite robust with a perfect score in all its occurrences. We consider these results as reasonable, although they might be improved by other means since the restraining force on the sole $\varepsilon$RMSD could be insufficient.

## Backbone conformations

The backmapping procedure applies a drift force uniquely on the bases, which are found to be in good agreement with the target structure. Sugar and phosphate groups, on the other side, will accommodate accordingly to the force field. In order to quantify the agreement of their geometry with the target structures, we analyze the glycosidic bond angle and the sugar pucker through the examination of the dihedrals $\delta$ and $\chi$, since they are directly related to the base orientation. The agreement of these quantities is shown in Figure 2, where a set of representative structures is presented. The figures corresponding to the rest of the structures is contained in the SM (Figs. 2-8). The nucleotides are colored in red if less than 30% of the structures agree with the native value, yellow if the agreement is between 30% and 70% and green otherwise.

We observe that the agreement is in general very good. Most of the disagreements are relegated to bulges, terminal and loop regions, as in 1CSL shown in Fig. 2a and 2b. It is also noticeable that when disagreements are present the sugar puckers are not reproduced as well as the glycosidic bond angle conformation. In some instances, as in the marked nucleotide in the fragment of 1XJR (Figure 2f), the native orientation of the base is obtained, but as a compensation of a wrong sugar pucker (C2' endo) and a distorted $\chi$ angle. The same happens in the sarcin-ricin loop of 1MSY (shown in SM), and in a fraction of the converged simulations in the G9 nucleotide of 2KOC, shown in yellow in Fig. 2h. Despite this, the correct orientation of the base with respect to the stem as in the native structure allows the formation of its native non-canonical interactions. On the contrary, 2KOC and 2GDI both possess bases which are flipped uniquely by virtue of the sugar puckers, remarkably reproduced in all the simulations and simultaneously reproduced with the right $\chi$ conformation. The quality of the backbone conformations is also of relevance for the base-backbone interactions, as observed in 1EVV, where a non-native sugar pucker is directly correlated with a missing base-phosphate contact. Some hairpin loops as 1ZIH and 2GDI reproduce almost perfectly the conformations of their native structures (see SM).

The analysis of the remaining rotamers that define the backbone conformation also shows good agreement and a similar trend, being specifically sensitive in the $\alpha$ and $\beta$ dihedral. These results are presented in the SM.

## Effect of ℰRMSD parameters and comparison with RMSD pulling

To conclude the analysis, we compare different sets of parameters, concerning the strength of the pulling force and the cutoff used in the ℰRMSD definition. We also test the same pulling procedure using the standard RMSD. For a direct comparison with ℰRMSD, we consider C2, C4 and C6 atoms only. The structures analyzed are the 2KOC hairpin and the 1L2X pseudoknot. We define seven different protocols: S0, S1, S2, S3, S4, S5 and S6, which are listed in Table 3.

The constant $\kappa_{max}$ for the RMSD was scaled from the ℰRMSD constant since both distances are linearly related for small fluctuations around the native structure (see SM).

The folding pathway can be investigated monitoring the ℰRMSD with respect to the native structure as a function of time. In S0, the stem was formed in only 2 of 10 initial conditions, while in the other cases there were no simulations with an extended strand as the final conformation. Fig 3a shows this quantity for representative simulations of the hairpin under the S0, S1, S2, S3 and S4 scenarios, together with the RMSD for S5 and S6 of the same system in Fig. 3b. We observed that in the favorable cases of S0 and in some realizations of S1, the strand stayed extended for a long time while some structure of the tetraloop started to develop, as shown in Fig. 3c, to continue with an abrupt decrease of the ℰRMSD which corresponded to the formation of the stem shown in Fig. 3d. In S1, we also observed a second case where the stem is formed at around 500 ps, followed by the formation of the tetraloop. A snapshot previous to the folding of this case is shown in Fig. 3e. For S2, the pulling force towards the native structure is stronger, and we observe the same two cases as in S1. In S3, we note that the stem is formed earlier in all the cases that converged, and later on the tetraloop is arranged. Finally, in the S4 system, with both a strong coupling and a larger cutoff, we observe a similar behavior as in S3, but at a much faster rate. In all cases, except S1, all the initial conditions converged, although the agreement of their sugar puckers and glycosidic bond angle conformations is slightly worse in the nucleotide G9, which in the native structure is in *syn* state. On the other hand, RMSD-driven S5 simulations were not able to converge and produce good results, since the closing pair of the nucleotide G9 was not properly oriented in any case. S6, on the other hand, converged pretty quickly, with a folding divided into clearly distinguishable steps, in the same manner of the second case of S1: that is, the stem formation preceded the arrangement of the loop, which in this case stood a long time closed before the guanine G9 could flip to its native position. Note that the steps of the folding are practically not visible from the sole RMSD examination.

For the pseudoknot, no initial condition folded in the case of S0, while three simulations folded successfully under the S1 parameters. In all these cases one hairpin is formed after the other (independent of the order), while four initial conditions could fold only one of them. When the strength of the restraint is increased, the folding procedure turns faster. In S2, four initial conditions converge to the native structure (Fig. 4a) while a misfolded structure is observed and two simulations are able to form only one hairpin. A similar success rate is found in S3; however, at least 5 initial conditions fold into a wrong structure, shown in Fig. 4b. This is also observed in S4, and in only one case of S2. Despite this, in S4 two initial conditions converge. For RMSD-driven S5 and S6 simulations, we observe a completely different scenario. Here, the strand collapses as a whole, where it slowly rearranges its bases. In most cases, however, the whole arrangement is good although in several cases the nucleotides C13, G15 of G17 are flipped with respect to their native counterpart. The INF for base pairs, stacking and base-phosphate interactions and the radius of gyration as functions of time are shown in Fig. 4c, 4d, 4e and 4f for S6. We see again the folding of S1 in two steps, while S6 goes continuously. Also, the abrupt fall of $R_G$ characterizes the folding process in comparison to the $\mathcal{E}$RMSD-driven folding.

For completeness of the analysis, the $\mathcal{E}$RMSD and some selected of snapshots during their folding pathway are illustrated in the SM (Figs. 9-13).

Finally, we took some of the misfolded structures obtained from the S5 protocol of both structures, and applied the S1 backmapping on them. In the case of 2KOC, the G9 nucleobase was flipped to its native orientation in 1 of 10 initial conditions, while the in the rest it went out of the loop without adopting its native position. For 1L2X, we took two misfolded initial conditions, with misplaced and flipped bases, and ran them over 5 initial conditions. We observe that the flipped bases always converge to their native orientation, but the position of some misplaced bases is not successfully corrected (see Fig. 14 in SM) under this protocol.

# Discussion

We have shown that the backmapping procedure studied here in general produces good results over a dozen of fragments with different complexity. The conformation of the backbone, and specifically the sugar pucker and glycosidic bond angles, are reconstructed in most cases, which stresses the importance of the base arrangement and their interaction network in the constitution of a RNA structure [23]. Nevertheless, bulges and terminal nucleotides seem to be more difficult to capture, which is not surprising due to their higher flexibility and possible additional interactions present in the crystal environment. In addition, they are situated in regions with a lower density of nucleotides compared to stems, for example. In this regard, following the definition of $\mathcal{E}$RMSD, the pulling force might be weaker due to the lower number of neighbors considered in their surrounding environment. The interaction network obtained also reproduces the native one in most cases, although the reconstruction of all the base-phosphate interactions might require the imposition of additional constraints on the phosphorous atoms.

It is advisable to keep in mind the limitations of the current approach. The quality of the results is expected to depend on the force field accuracy, and the binding of ions to sites deeply embedded in the structure (as it happens for instance in the case of DNA quadruplexes [49]) will probably require of slower pulling rates to allow a better exploration of the configurational space.

With respect to the use of other parameters and cutoffs for the $\mathcal{E}$RMSD, it seems that the choice proposed here as S1 is reasonable and it allows to explore multiple conformations towards the folding of the native structure, as shown in the UUCG tetraloop and the studied pseudoknot. RMSD pulling can be a safe choice for obtaining the right global structure, with a reasonable interaction network. However, local details such as the orientations of the bases might be suffocated by the agreement of the structure on a larger scale, and therefore, lead to wrong results difficult to detect without the monitoring of the $\mathcal{E}$RMSD. A possible solution is to using a start from an atomistic structure that is closer to the target. In the present study, however, we found that a sequential application of the RMSD and $\mathcal{E}$RMSD minimization protocols might not be very efficient. It is therefore recommendable to perform several simulations starting from unbiased initial conditions, although this might reduce the efficiency of the method and limit it to structures with a reduced number of nucleotides, as the ones studied here. On the other hand, $\mathcal{E}$RMSD pulling starting from a random coil could generate folding pathways more realistic and contribute to a better understanding of the folding process [50]. It is interesting to observe that in a very recent paper $\mathcal{E}$RMSD and RMSD from native were simultaneously used in order to drive folding of a DNA quadruplex [49].

# Acknowledgements

This work has received funding form the European Research Council under the European Union's Seventh Framework Programme (FP/2007-2013)/ ERC Grant Agreement n. 306662, S-RNA-S.

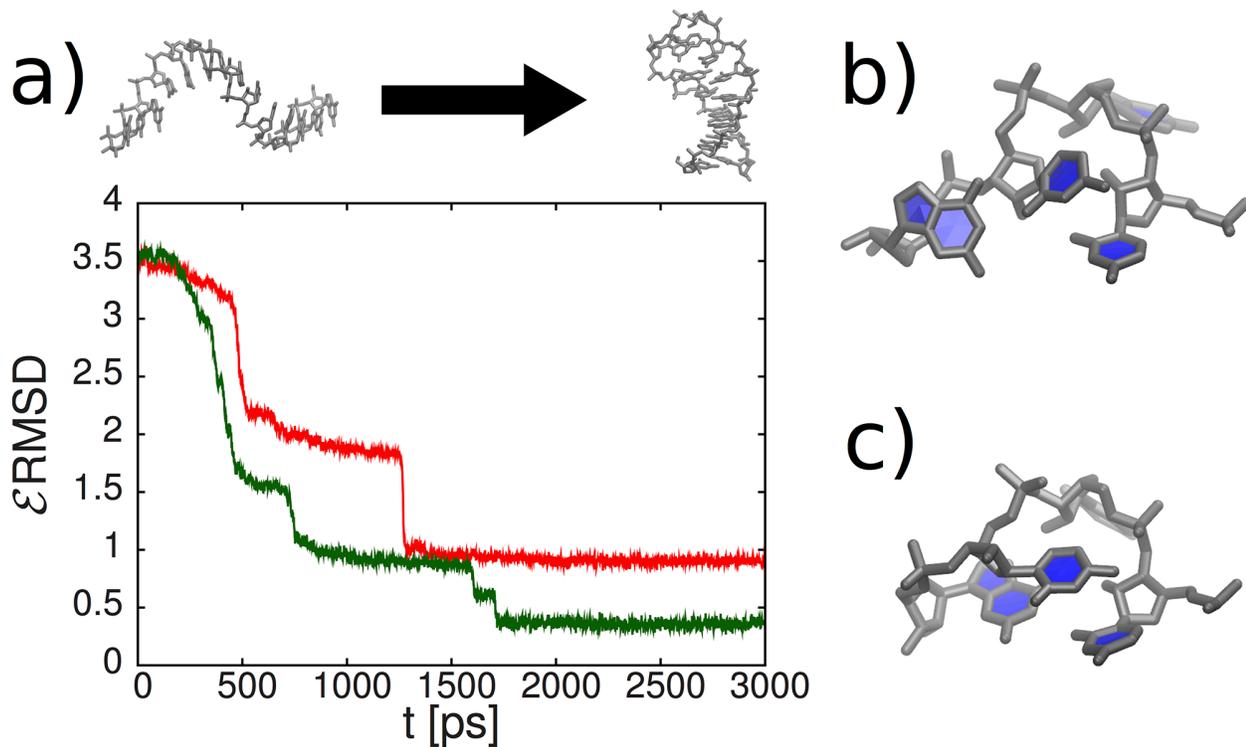

Figure 1: a) Two independent simulations for the backmapping of the UUCG hairpin. Green curve converges, red one gets trapped in local minimum. b) Misfolded loop. c) Folded loop.

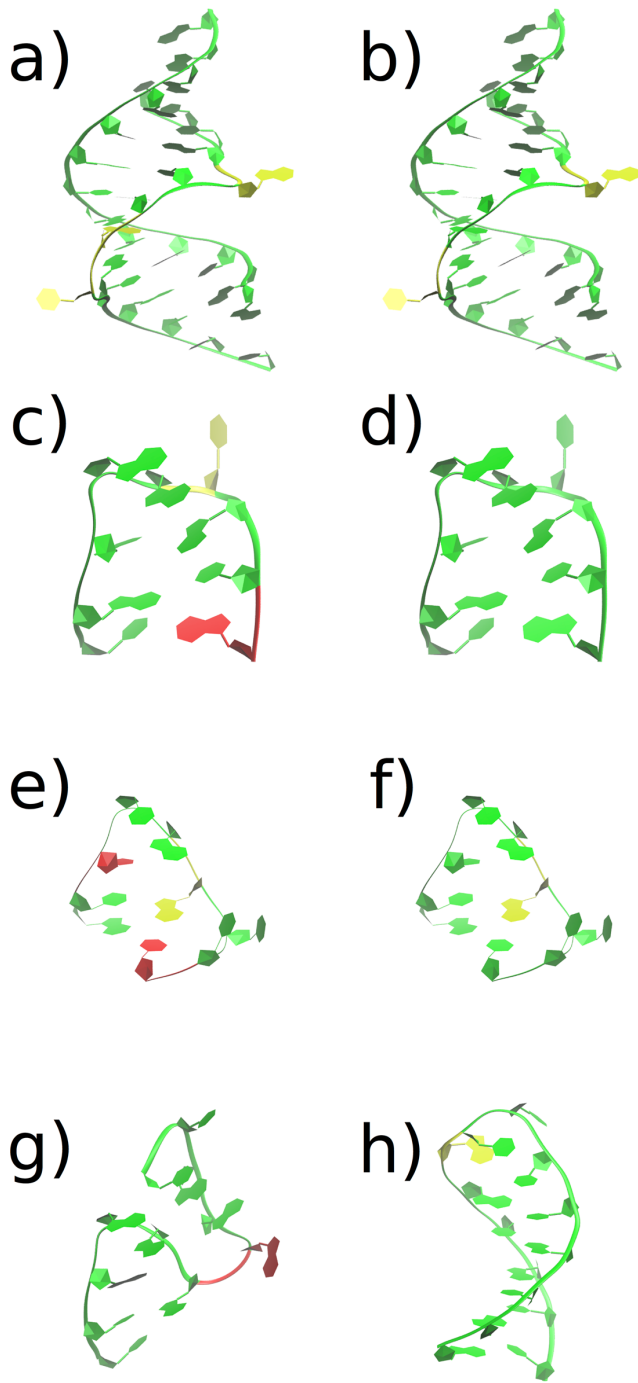

Figure 2: Depiction of consistency of sugar puckers and glycosidic bond angle conformations for selected structures. Nucleotides are colored in green when the agreement is larger than 70%, yellow when it is between 30% and 70% and red otherwise. a) and b) show 1CSL, for sugar puckers and $\chi$, c) and d) show 1XJR and e) and f), 1EVV, in the same order. In the other two cases, there is no difference between the pucker and $\chi$ figures, so only one represents 2CKY in g) and 2KOC in h).

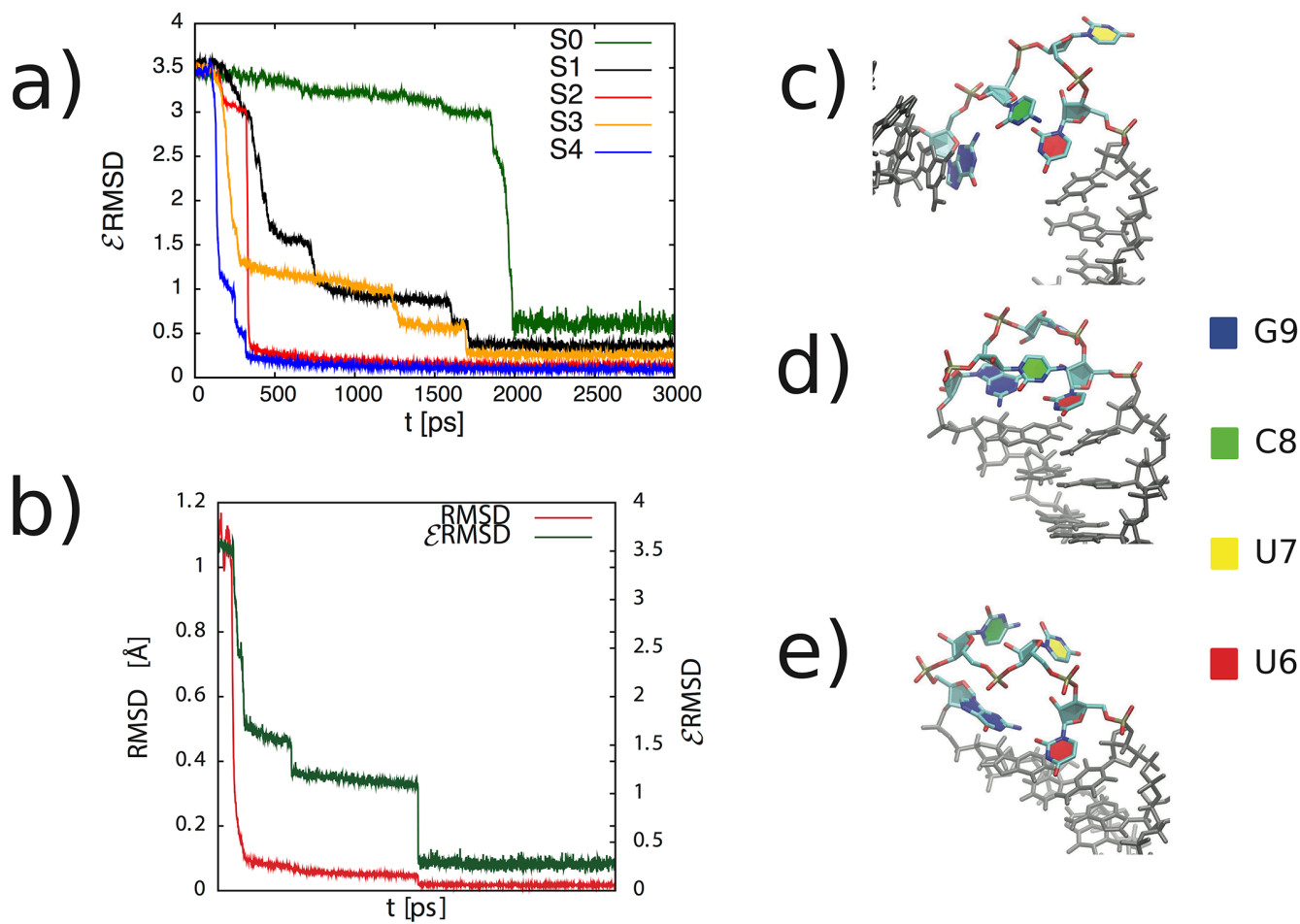

Figure 3: Results for the pulling simulations of the UUCG tetraloop. a) $\mathcal{E}$RMSD as a function of time for the $\mathcal{E}$RMSD pulled systems. The curve of S1 corresponds to the second case described in the text, while in S2 we plotted the first case, which is similar to S0. b) RMSD and $\mathcal{E}$RMSD as a function of time for the RMSD pulled system S6. S5 showed no convergence. c) shows a snapshot of the loop in S1 before closing the stem and converging to the native structure depicted in d), while e) shows the loop when the stem closes before its arrangement, as observed also in S1 and in the rest of the simulations.

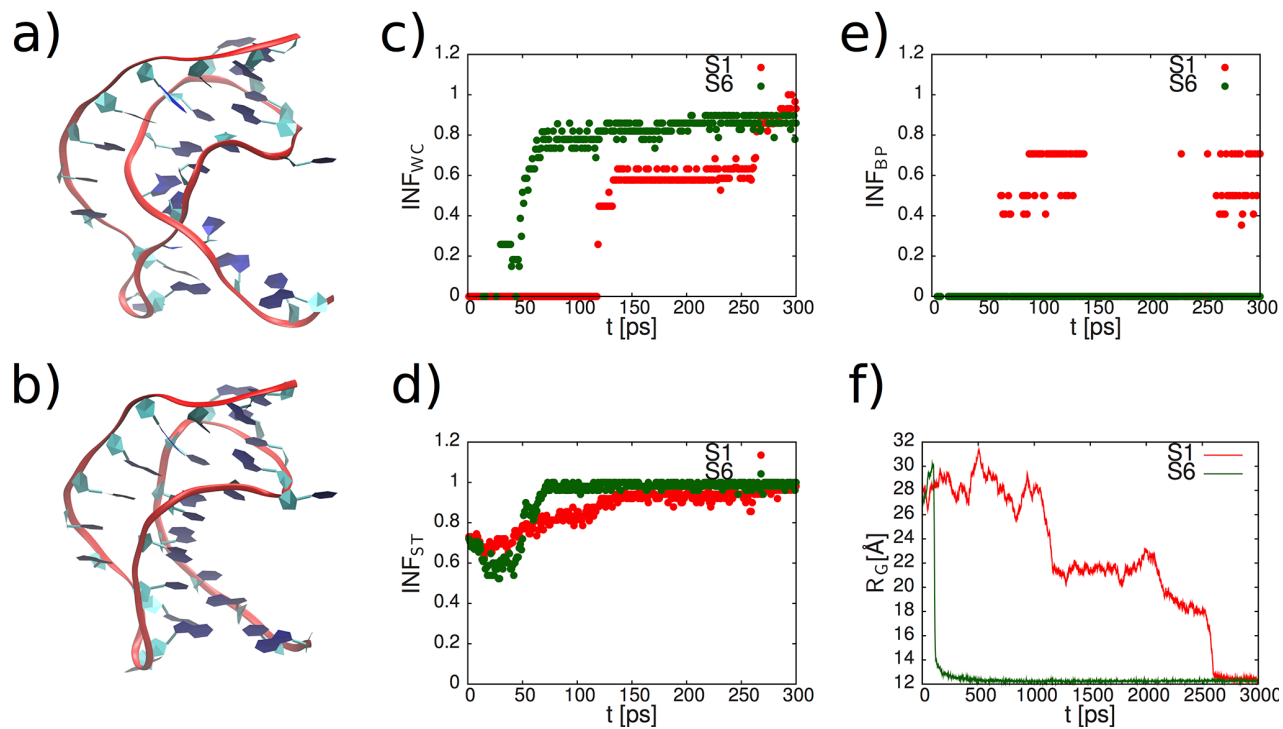

Figure 4: Results for the pulling simulations of the pseudoknot. a) Folded structure from S1, b) misfolded structure from S4. INF for base-pair, stacking and base-phosphate interactions are in c), d) and e). Radius of gyration as a function of time for converged simulations is in f).

| Structure | N | Fragment | Resolution (Å) | Description |
|---|---|---|---|---|
| 2CKY | 11 | A26-A36 | 2.9 | Single strand |
| 1CSL | 28 | Full | 1.6 | Duplex |
| 1I9X | 26 | Full | 2.18 | Duplex |
| 1L2X | 27 | Full | 1.25 | Pseudoknot |
| 1MSY | 27 | Full | 1.41 | Hairpin; Sarcin-Ricin domain |
| 1XJR | 9 | A20-A28 | 2.7 | Hairpin |
| 1ZIH | 12 | Full | - | Hairpin (GCAA tetraloop) |
| 2GDI | 11 | X25-X35 | 2.05 | Hairpin |
| 2KOC | 14 | Full | - | Hairpin (UUCG tetraloop) |
| 2LA5 | 27 | Full | - | Quadruplex |
| 2OIU | 22 | P20-P41 | 2.6 | Duplex; large stem |
| 1EVV | 9 | A53-A61 | 2.0 | Hairpin |

Table 1: List of RNA fragments studied with their description. N denotes the number of nucleotides. Resolution is included for X-ray structures.

| Structure | $N_C$ | $INF_{st}$ | $INF_{wc}$ | $INF_{nwc}$ | $INF_{bph}$ |
|---|---|---|---|---|---|
| 2CKY | 3 | 0.93 | 1 | 1 | 0 |
| 1CSL | 10 | 0.96 | 1 | 1 | 0.9 |
| 1I9X | 10 | 0.97 | 1 | - | - |
| 1L2X | 3 | 0.97 | 0.87 | 0.7 | 0.33 |
| 1MSY | 4 | 0.94 | 0.99 | 1 | 0.18 |
| 1XJR | 10 | 0.94 | 1 | 1 | 0.71 |
| 1ZIH | 10 | 1 | 0.98 | 0.9 (1) | 0.3 |
| 2GDI | 10 | 1 | 0.93 | 0.5 (1) | - |
| 2KOC | 5 | 0.96 | 1 | 1 | 1 |
| 2LA5 | 1 | 0.92 | 0.86 | 0.82 | 0 |
| 2OIU | 1 | 1 | 0.91 | 1 | 0 |
| 1EVV | 6 | 0.98 | 1 | 1 | 0.74 |

Table 2: Number of initial conditions which converged to the native structure, $N_c$, from a total of 10. INF score for stacking, base pair, non-canonical base pairs and base-phosphate interactions, averaged over the minimum $\mathcal{E}$RMSD structure of each converged trajectory. The score is omitted when there are no interactions both in the native and simulated structures. In the case of 1ZIH and 2GDI, the number in parenthesis indicates the INF calculated considering the "nearly" formed contacts.

| Set | Pulling variable | $\varkappa_{max}$ | $\mathcal{D}_c$ |
|---|---|---|---|
| S0 | $\mathcal{E}$RMSD | 1000 kJ | 2.4 |
| S1 | $\mathcal{E}$RMSD | 1000 kJ | 3.6 |
| S2 | $\mathcal{E}$RMSD | 10000 kJ | 3.6 |
| S3 | $\mathcal{E}$RMSD | 1000 kJ | 6 |
| S4 | $\mathcal{E}$RMSD | 10000 kJ | 6 |
| S5 | RMSD | 60000 kJ/nm$^2$ | - |
| S6 | RMSD | 600000 kJ/nm$^2$ | - |

Table 3: Simulation protocols for UUCG tetraloop and 1L2X pseudoknot.

# Supplementary Material

## Rotamers

Table 1 contains the fraction of consistent structures with the native results for the dihedrals $\alpha$, $\beta, \gamma, \epsilon$ and $\zeta$. We have excluded from the analysis the stem regions, previously identified using the DSSR package [1,2]. These regions show a very good agreement in all the angles in general. The agreement was calculated in the same way as for the sugar pucker and glycosidic bond angle. In the remaining loops, we see that there is in general a good agreement over all the angles, which means that the backbone finds a way to reach its native conformation. In average, however, angles $\alpha$ and $\beta$ shows the worst fidelity towards the native structure. As expected, the worst cases are found in the bulges, in 1I9X, in the pseudoknot 1L2X outside of the stem, as well as in the loop regions of 1XJR. 2LA5 shows also low convergence in general, which is also reflected in this set of data.

| Structure | $\alpha$ | $\beta$ | $\gamma$ | $\epsilon$ | $\zeta$ |
|---|---|---|---|---|---|
| 2CKY | 0.4 | 0.48 | 0.82 | 0.97 | 0.78 |
| 1CSL | 0.83 | 0.72 | 0.88 | 1 | 0.83 |
| 1I9X | 0.68 | 0.4 | 0.83 | 1 | 0.31 |
| 1L2X | 0.24 | 0.5 | 0.33 | 0.97 | 0.6 |
| 1MSY | 0.65 | 0.61 | 0.88 | 1 | 0.8 |
| 1XJR | 0.57 | 0.43 | 0.99 | 1 | 0.82 |
| 1ZIH | 0.69 | 0.79 | 0.76 | 1 | 0.95 |
| 2GDI | 0.91 | 0.79 | 0.96 | 1 | 0.95 |
| 2KOC | 0.55 | 0.9 | 0.8 | 0.95 | 0.94 |
| 2LA5 | 0.34 | 0.25 | 0.3 | 1 | 0.66 |
| 2OIU | 0.46 | 0.38 | 0.63 | 0.99 | 0.69 |
| 1EVV | 0.7 | 0.85 | 0.83 | 1 | 0.78 |

Table 1: Average fraction of consistent dihedrals taken over all the nucleotides not forming canonical stems of selected structures.

# Numerical correspondence between $\mathcal{E}$RMSD and RMSD.

For the equivalence between RMSD and $\mathcal{E}$RMSD we found that $RMSD \approx C \times \mathcal{E}RMSD$, where $C = 0.13$ nm for both the hairpin and pseudoknot. A fit is shown in Fig. 1.

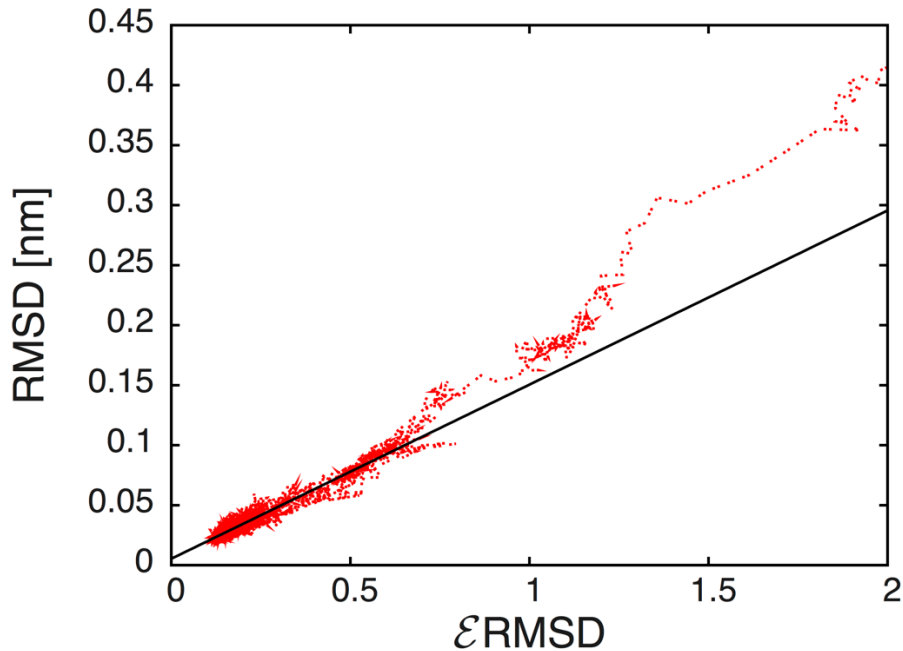

Figure 1: Fit of RMSD as a function of ERMSD from UUCG tetraloop

# Sugar puckers and glycosidic bond angle conformation of remaining structures.

Here we present the figures of the rest of the analyzed structures, colored according to the agreement of conformations of sugar pucker and glycosidic bond angle as described in the main text.

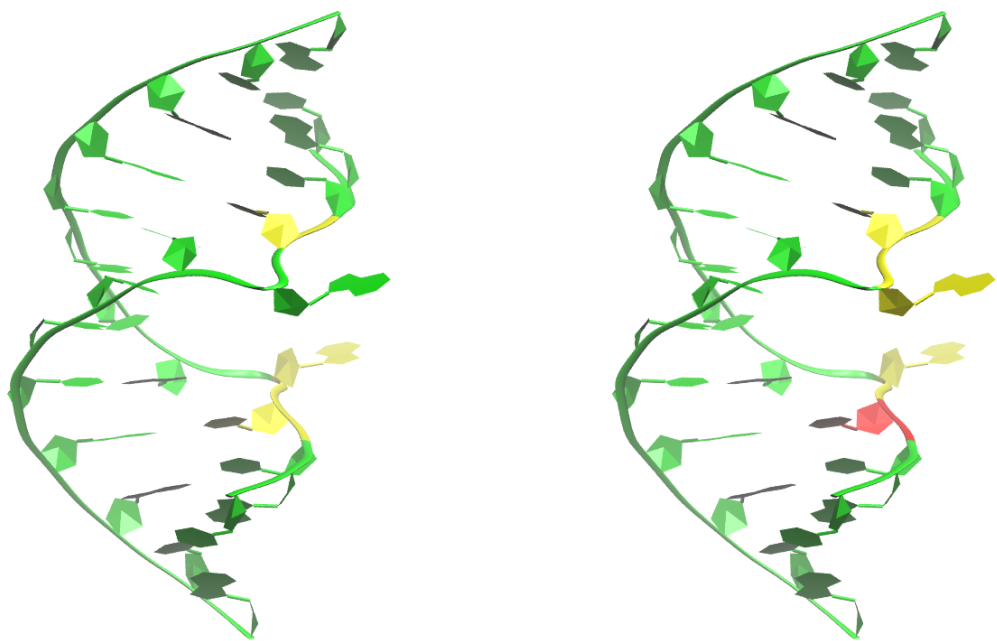

*Figure 2: 1i9x glycosidic bond angle conformation(left) and sugar puckers (right).*

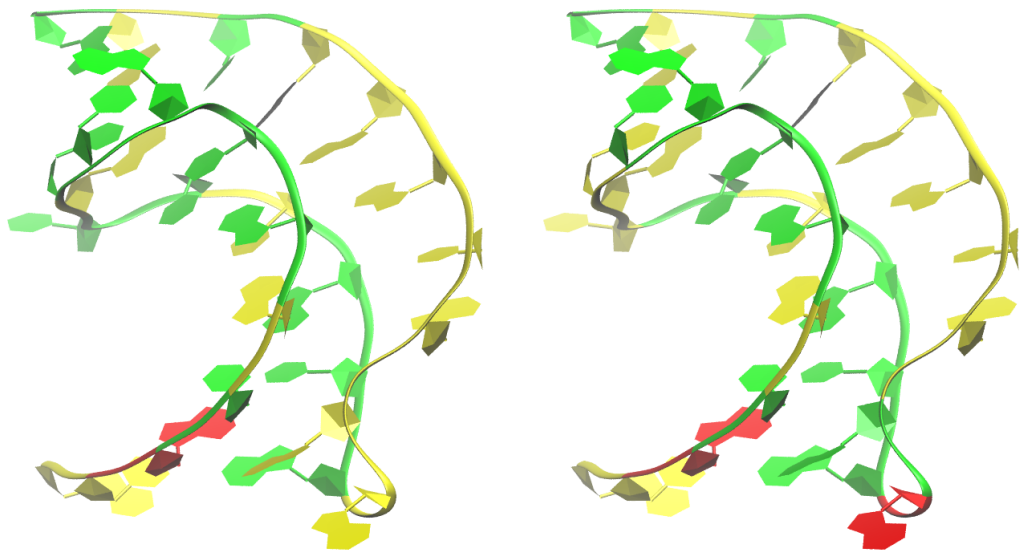

*Figure 3: 1l2x glycosidic bond angle conformations (left) and sugar puckers (right).*

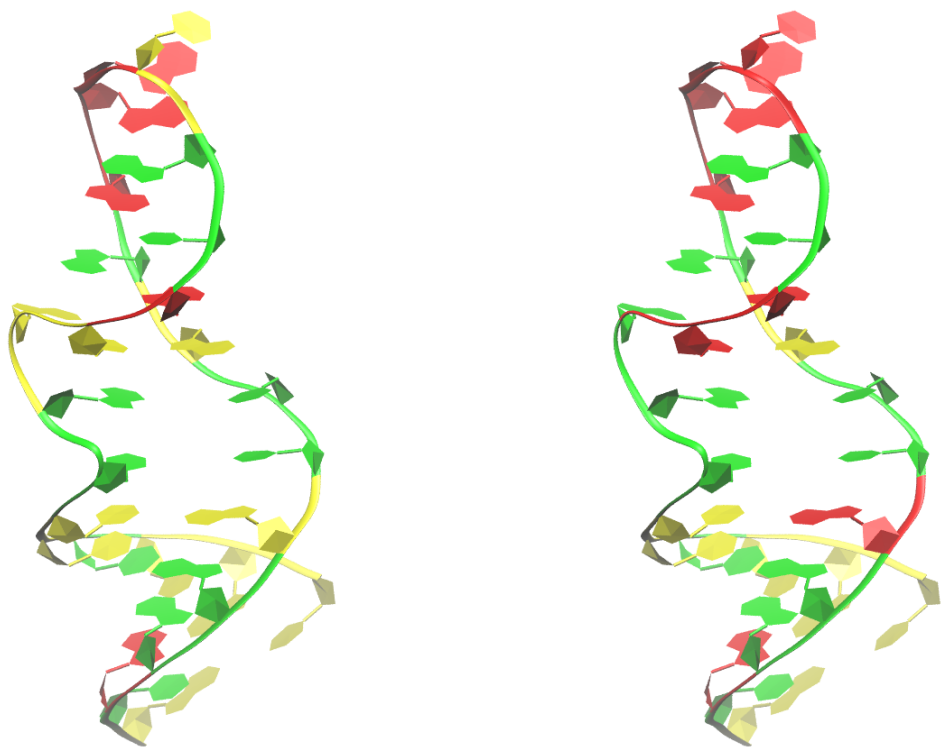

*Figure 4: 1msy glycosidic bond angle conformations (left) and sugar puckers (right).*

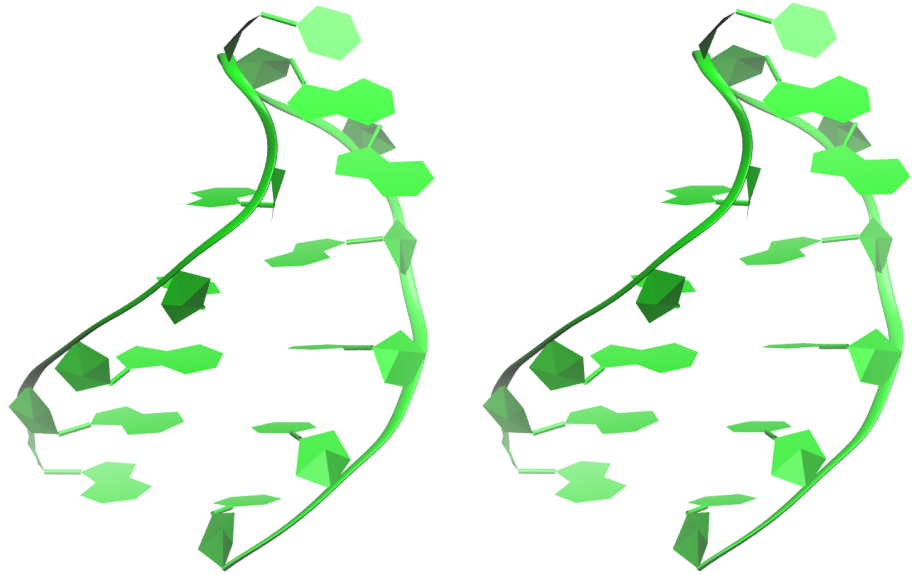

*Figure 5: 1zih glycosidic bond angle conformations (left) and sugar puckers (right).*

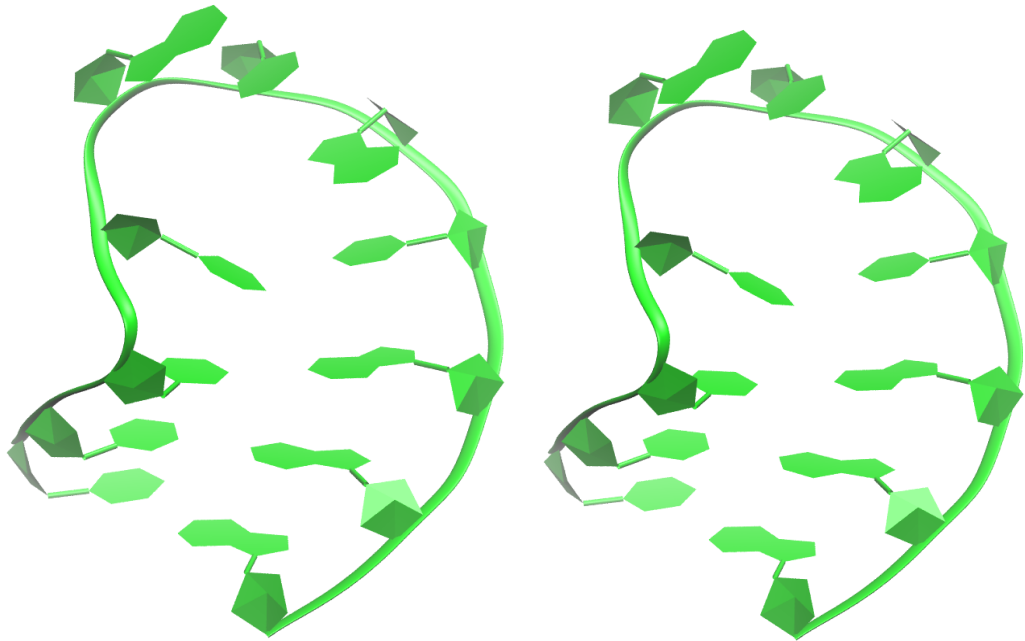

*Figure 6: 2gdi glycosidic bond angle conformations (left) and sugar puckers (right).*

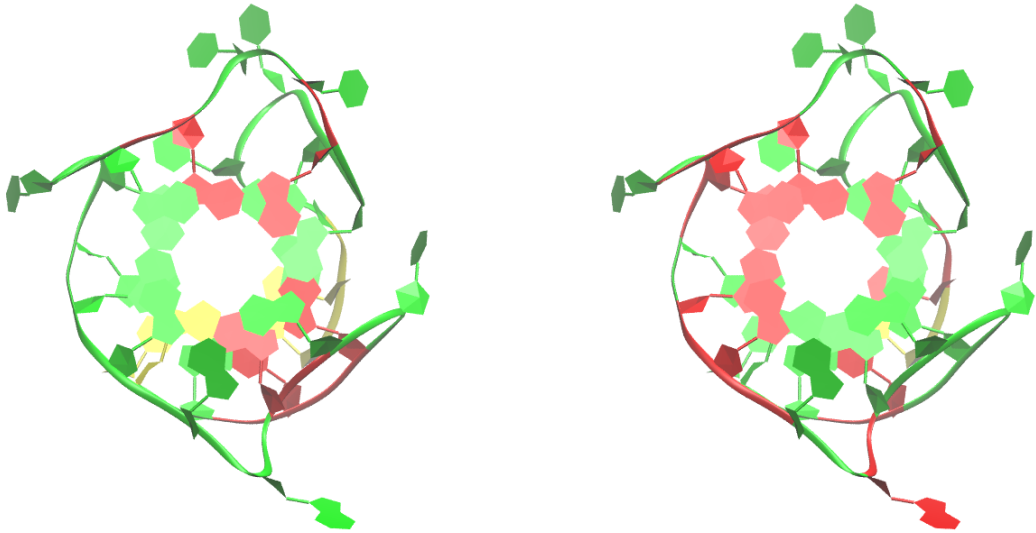

*Figure 7: 2la5 glycosidic bond angle conformations (left) and sugar puckers (right).*

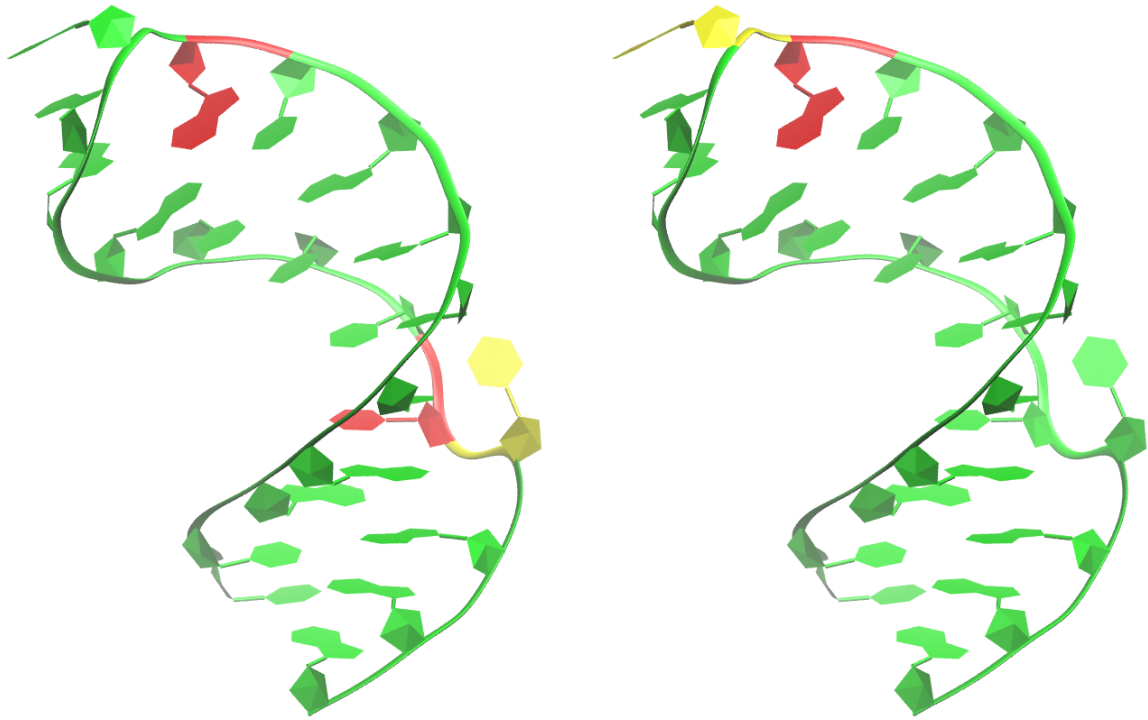

*Figure 8: 2oiu glycosidic bond angle conformations (left) and sugar puckers (right).*

# ERMSD and snapshots of folding pathway.

Here we depict the ERMSD of 1ZIH, 2LA5, 1MSY and 1L2X, together with snapshots of particular times. In addition, for 1L2X we include the same figures for the RMSD pulling.

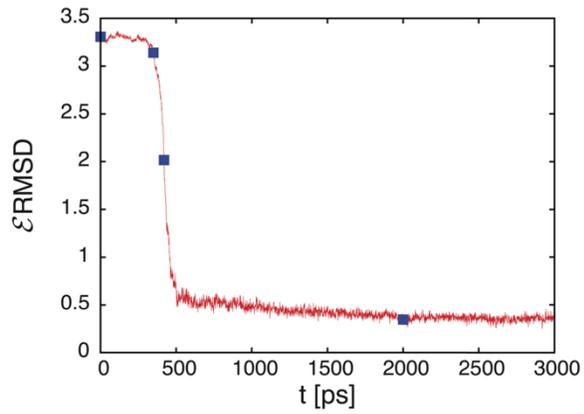 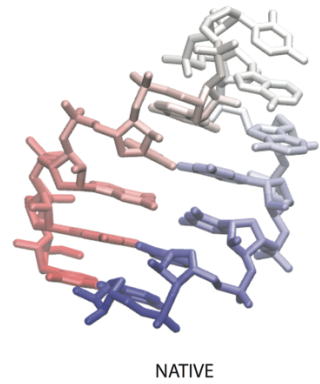

NATIVE

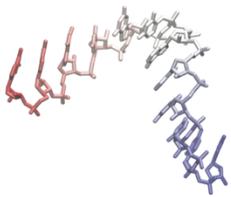 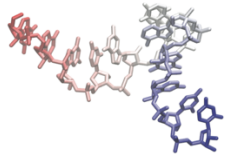 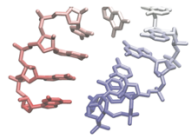 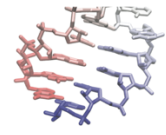

t= 0 ps     t= 350 ps     t= 420 ps     t= 2000 ps

*Figure 9: Pulling of 1ZIH*

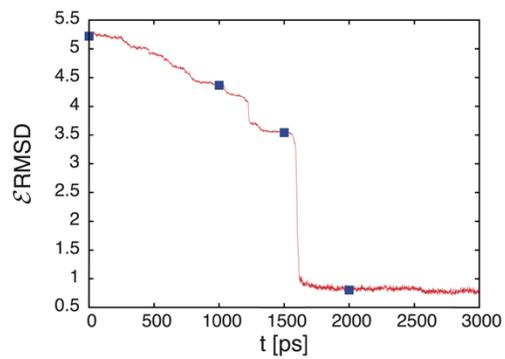
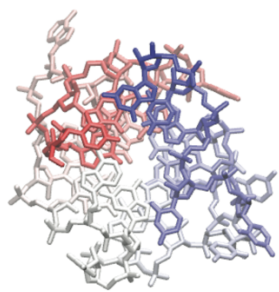
NATIVE

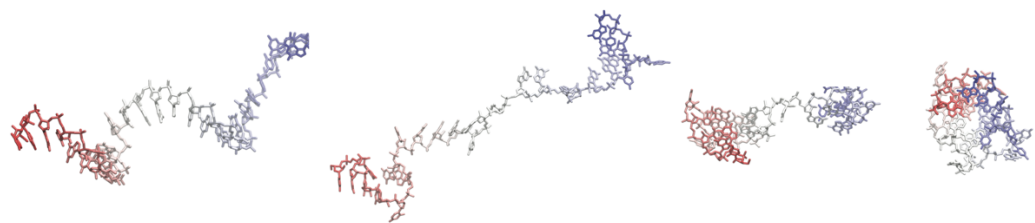

t= 0 ps      t= 1000 ps      t= 1500 ps      t= 2000 ps

*Figure 10: Pulling of 2LA5*

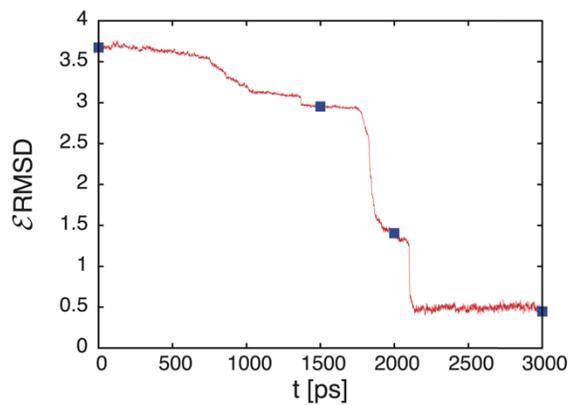 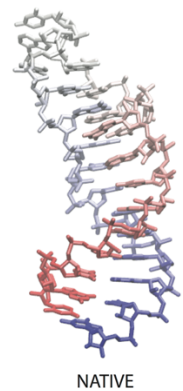

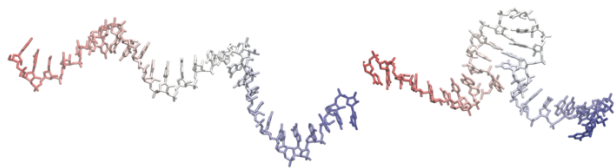 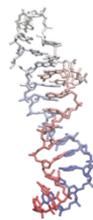 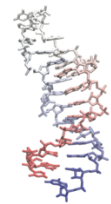

*Figure 11: Pulling of 1MSY*

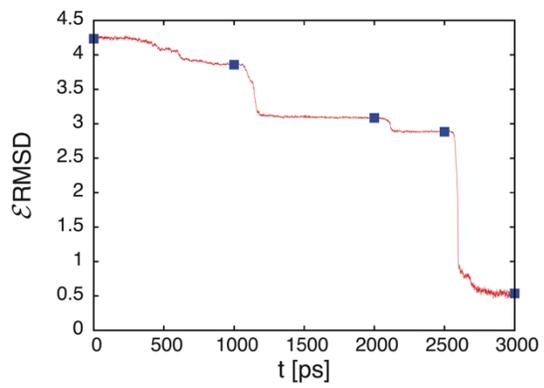
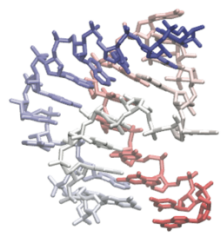

NATIVE

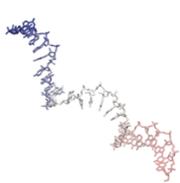
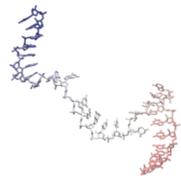
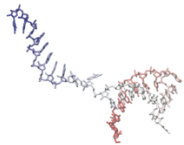
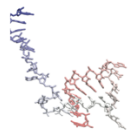

t= 0 ps    t= 1000 ps    t= 2000 ps    t= 2500 ps    t= 3000 ps

*Figure 12: Pulling of 1L2X*

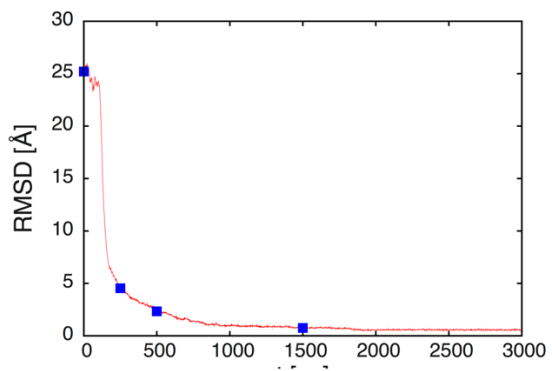
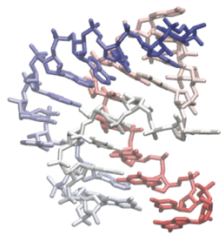

NATIVE

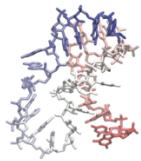
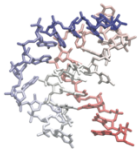
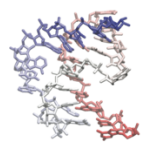

t= 250 ps			t= 500 ps			t= 1500 ps

*Figure 13: Pulling of 1L2X, RMSD*

# Loop inside 1L2X pseudoknot

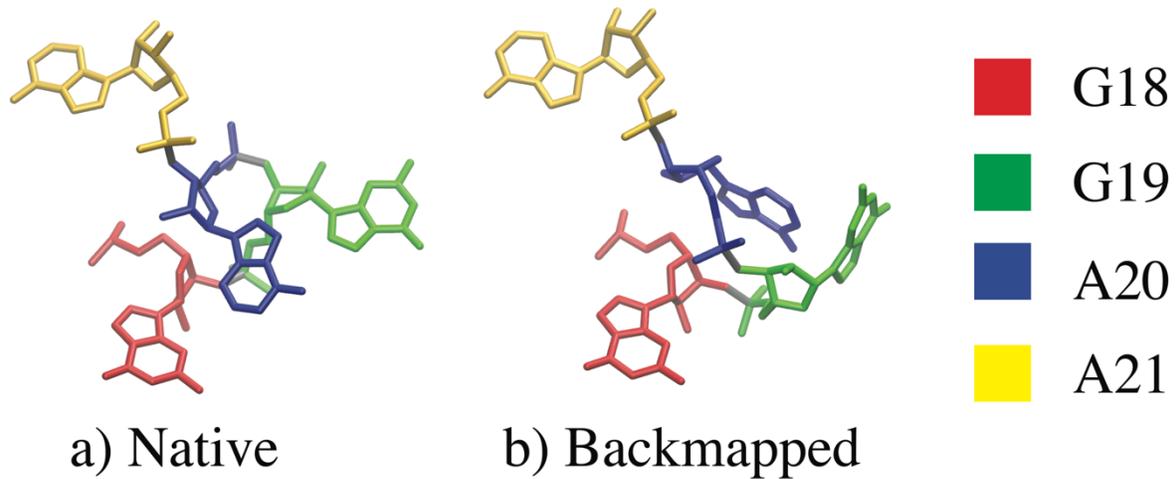

*Figure 14: Native and misfolded loop of 1L2X. The misfolded conformation was obtained after RMSD pulling and was not corrected by the subsequent ERMSD pulling.*